# Room-temperature ferromagnetism in the mixtures of the TiO$_2$ and Co$_3$O$_4$ powders

A. Serrano[1], E. Fernandez Pinel[2], A. Quesada[2,3], I. Lorite[4], M. Plaza[1], L. Pérez[1], F. Jiménez-Villacorta[5], J. de la Venta[1], M. S. Martín-González[6], J. L. Costa-Krämer[6], J. F. Fernandez[4], J. Llopis[1], M. A. García[1,4*].

[1]Dpto. Física de Materiales Universidad Complutense de Madrid, Madrid, Spain
[2]Instituto de Magnetismo Aplicado UCM, Las Rozas, Madrid, Spain.
[3]Instituto de Ciencia de Materiales de Madrid, CSIC, Madrid, Spain
[4]Instituto de Cerámica y Vidrio, CSIC, Madrid, Spain
[5]SpLine, Spanish CRG beamline at the ESRF, Grenoble, France and Instituto de Ciencia de Materiales, CSIC, Madrid,Spain.
[6]Instituto de Microelectrónica de Madrid, CSIC, Tres Cantos, Madrid, Spain

*Corresponding author: email: ma.garcia@fis.ucm.es

## Abstract

We report here the observation of ferromagnetism (FM) at 300 K in mixtures of TiO$_2$ and Co$_3$O$_4$ powders despite the antiferromagnetic and diamagnetic character of both oxides respectively. The ferromagnetic behavior is found in the early stages of reaction and only for TiO$_2$ in anatase structure; no FM is found for identical samples prepared with rutile-TiO$_2$ . Optical spectroscopy and X-ray absorption spectra confirm a surface reduction of octahedral Co$^{+3}$→Co$^{+2}$ in the mixtures which is in the origin of the observed magnetism.

## Introduction

The appearance and control of magnetism in traditionally non-magnetic oxides is nowadays one of the most active and pursued goals of material physics [1]. In the last years, research has been focused mainly on oxides doped with magnetic ions (the so-called Diluted Magnetic semiconductors). Recent results [2,3,4,5] indicate that the appearance of magnetism in these oxides (mainly ZnO, and $TiO_2$) is related to the presence of defects and surface/interface effects [6,7] but the origin of most of the experimental results is still unclear [8]. Actually, results are hardly reproducible and findings from different groups are commonly contradictory. Understanding this lack of reproducibility is probably one of the main challenges to determine the origin of this magnetism.

A common feature of all the experimental observations of this magnetism is that signals are very weak, suggesting that only few atoms are involved in the magnetic response. Thus, it has been proposed that the emerging magnetism in oxides corresponds to surface/interface magnetism [9,10,11].

It is really tough understanding the origin of this magnetism based only on magnetic measurements, since signals are very weak and effects from the rest of the material as impurities could mask the signals coming from interfaces. In this framework, correlating the magnetic properties with other measurements, sensitive to the electronic structure, can help to clarify the origin of this magnetism. This is the case of optical measurements, as the energy of photons involved in the processes allows discriminating between different processes produced in different atoms. Furthermore, optical properties are particularly sensitive to surfaces as the broken symmetry of the surface induces new electronic levels (surface states), so they can be especially useful for the investigation of surface effects.

In this work we study the magnetic properties of $TiO_2$ and $Co_3O_4$. Both oxides are mixed in powder form and thermal treated to promote their interaction. Optical and X-ray absorption measurements are used to identify the origin of the observed room temperature magnetism. The difficulties to reproduce experiments in this kind of materials are also addressed.

**Experimental**

Samples were prepared by mixing powders of $TiO_2$ in anatase (A-$TiO_2$) and rutile structure (R-$TiO_2$) with 1% and 5% wt $Co_3O_4$. Analitycal grade powders were selected with average particles size in the submicronic range, typically 0.3-0.4 µm (see figure 1). Pure powders were tested in order to ensure that no ferromagnetic contributions were present. $TiO_2$ powder showed diamagnetic behavior and $Co_3O_4$ paramagnetic one. When subtracting the paramagnetic behavior, some $Co_3O_4$ samples exhibited a weak ferromagnetic like contribution with magnetization below $5 \cdot 10^{-4}$ emu/g that disappeared after annealing at 400°C [11]. This residual magnetization due to trace impurities was taking into account and adequately subtracted from the magnetic measurement of mixtures. Any powder with higher ferromagnetic-like behavior was considered as contaminated and completely discarded. Initially, the powders were mixed and milled in a attrition mill with zirconia balls for 15 minutes and subsequently they were thermally treated in air during 12 hours at different temperatures between 500°C and 900 °C. Selected raw materials were also processed following the same procedure to verify experimental conditions. Three different sets of samples were independently prepared at different laboratories (corresponding to authors affiliations 1, 2 and 3). Processing was managed without using any kind of metallic tools to prevent contamination

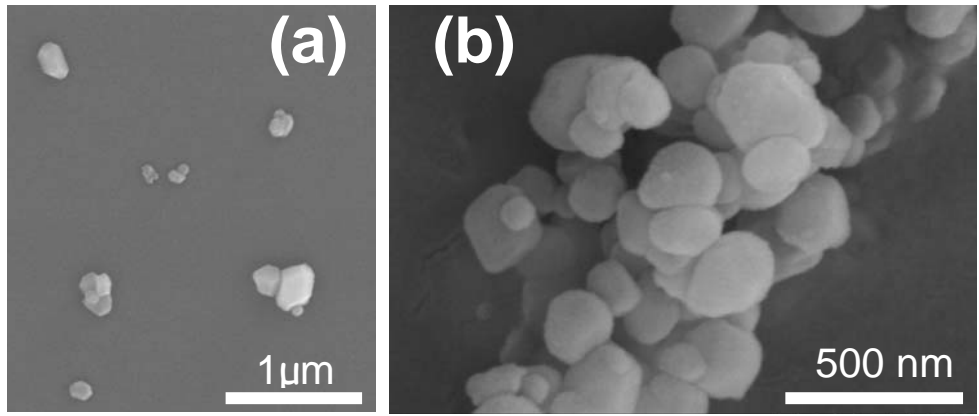

**Figure 1.** SEM images from (a) Initial $Co_3O_4$ powder. (b) 99%A-$TiO_2$-1%$Co_3O_4$ sample milled.

The structural analysis of the samples was carried out with a Siemens D5000 X-Ray Diffractometer using a monochromatic Cu K line and operating at 40 kV and 40 mA. Magnetic characterization was performed in three different Vibrating Sample Magnetometers: a VSM LDJ Instruments, a Quantum Design PPMS-VSM and a VSM Lakeshore 7304. For the magnetic measurements, all possible sources of experimental errors described in [12] were taken into account. Optical absorption was measured with a Shimadzu 3101 spectrophotometer attached with an integrating sphere. X-ray Absorption Spectroscopy was performed at beamline BM25 (SpLine) in the European Synchrotron Radiation Facility at Grenoble.

## Results and discussion

Figure 2 presents XRD measurements for the samples prepared using R-$TiO_2$. Up to annealing temperatures of 600°C only the initial R-$TiO_2$ (R) and $Co_3O_4$ phases are observed in the XRD patterns. After annealing at 700°C, some peaks from $CoTiO_3$ appear for the samples with 5%$Co_3O_4$, coexisting with the initial phases. These data confirm that the reaction between both phases has started at this temperature. The peaks are scarcely identified for the samples with 1% $Co_3O_4$. This is quite reasonable since the amount of $CoTiO_3$ that can be formed is limited mainly by the concentration of $Co_3O_4$ in the sample. Thus, the differences between both samples seem more related to the detection limit of the experimental setup than to structural differences. As the temperature increases over 700°C, the peaks from CoTiO$_3$ also do, while the Co$_3$O$_4$ ones decrease, disappearing at 800°C. At this temperature, all the Co$_3$O$_4$ in the sample has reacted with the TiO$_2$ forming CoTiO$_3$. Due to the small fraction of Co$_3$O$_4$ there is an excess of TiO$_2$ that remains unaltered in the sample at all annealing temperatures, as evidenced in the XRD patterns.

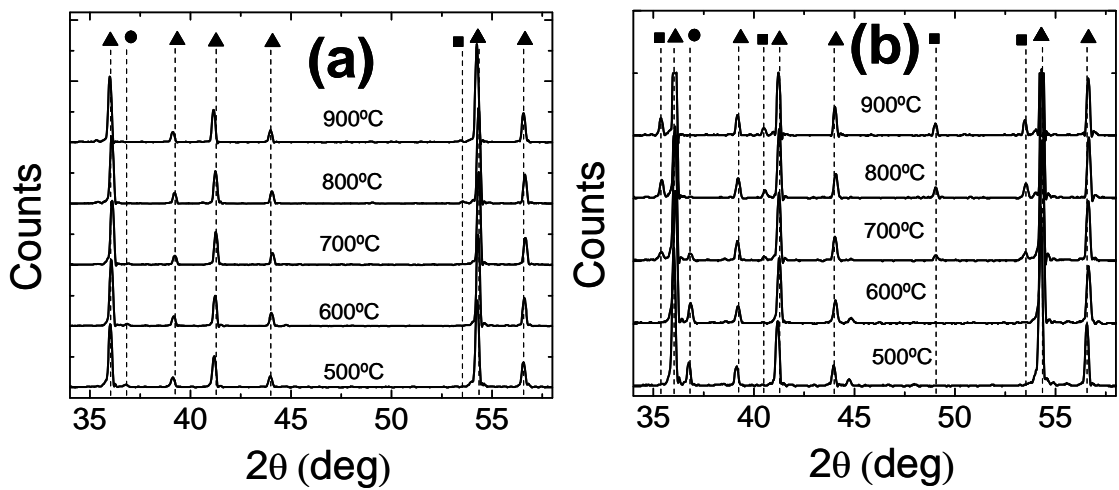

**Figure 2**. XRD patterns from R-TiO$_2$ with (a) 1% mol of Co$_3$O$_4$ and (a) 5% mol of Co$_3$O$_4$. Symbols stand for (▲) TiO$_2$ rutile, (●) Co$_3$O$_4$ and (■) CoTiO$_3$.

Summarizing, the annealing process induces the interaction between both phases, consisting in the reaction of the Co$_3$O$_4$ with a small amount of R-TiO$_2$ to form CoTiO$_3$. The reaction starts below 700°C and is completed over 800°, when all the Co$_3$O$_4$ has reacted to form the spinel phase that coexists with the excess of R-TiO$_2$.

The XRD patterns from the samples prepared with A-TiO$_2$ are more complicated (figure 3). As in the case of R-TiO$_2$, up to 600°C only the initial phases are observed in the XRD diffractograms. After annealing at 700°C the presence of scarce peaks, ascribed to CoTiO$_3$, coexisting with the initial phases, indicates a partial reaction. Simultaneously, peaks from R-TiO$_2$ are shown, confirming that the transformation of A-TiO₂→R-TiO₂ has started at this temperature. The presence of CoTiO₃ increases with the temperature of the thermal treatment, while the transformation of A-TiO₂→R-TiO₂ is completed at 900°C. At this temperature, only R-TiO₂ and CoTiO₃ are present in the sample (as for the samples prepared with R-TiO₂).

Therefore, for these samples, the annealing produces two simultaneous processes: The reaction of TiO₂ with Co₃O₄ (similar to that produced in samples with R-TiO₂) and the transformation A-TiO₂→R-TiO₂.

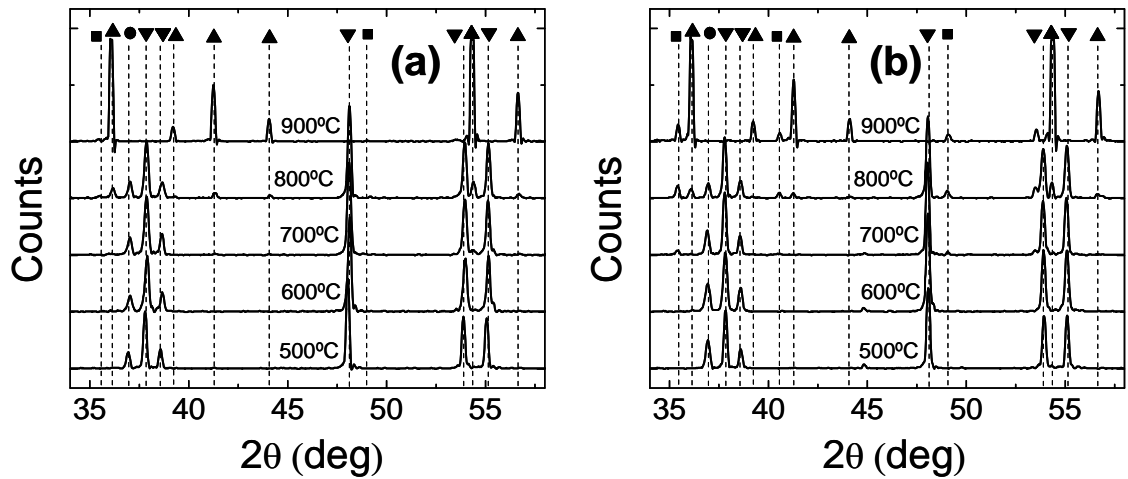

**Figure 3**. XRD patterns from A-TiO₂ with (a) 1% mol of Co3O4 and (a) 5% mol of Co₃O₄. Symbols stand for (▼) TiO₂ anatase, (▲) TiO₂ rutile, (●) Co₃O₄, (■) CoTiO₃.

Magnetic characterization of the samples is presented in figure 4. The overall behavior is paramagnetic with susceptibility decreasing up to annealing temperature of 500°C and increasing for higher temperatures. According to XRD diffractograms, the increase in magnetic susceptibility is due to the reaction of the Co₃O₄ with the TiO₂ to form CoTiO₃ that has larger magnetic susceptibility than Co₃O₄ [13,14]. Differences in the value of magnetic susceptibility for rutile and anatase samples can be explained in terms of the different kinetic of the reaction with Co₃O₄. Rutile exhibits a more closed structure than anatase [15], so the reaction with Co₃O₄ is slower than for anatase, that exhibits a more opened structure and is therefore more reactive. Hence, for a fixed temperature, the degree of transformation is smaller for rutile, and the magnetic susceptibility decreases accordingly (as the transformed $CoTiO_3$ exhibits larger susceptibility).

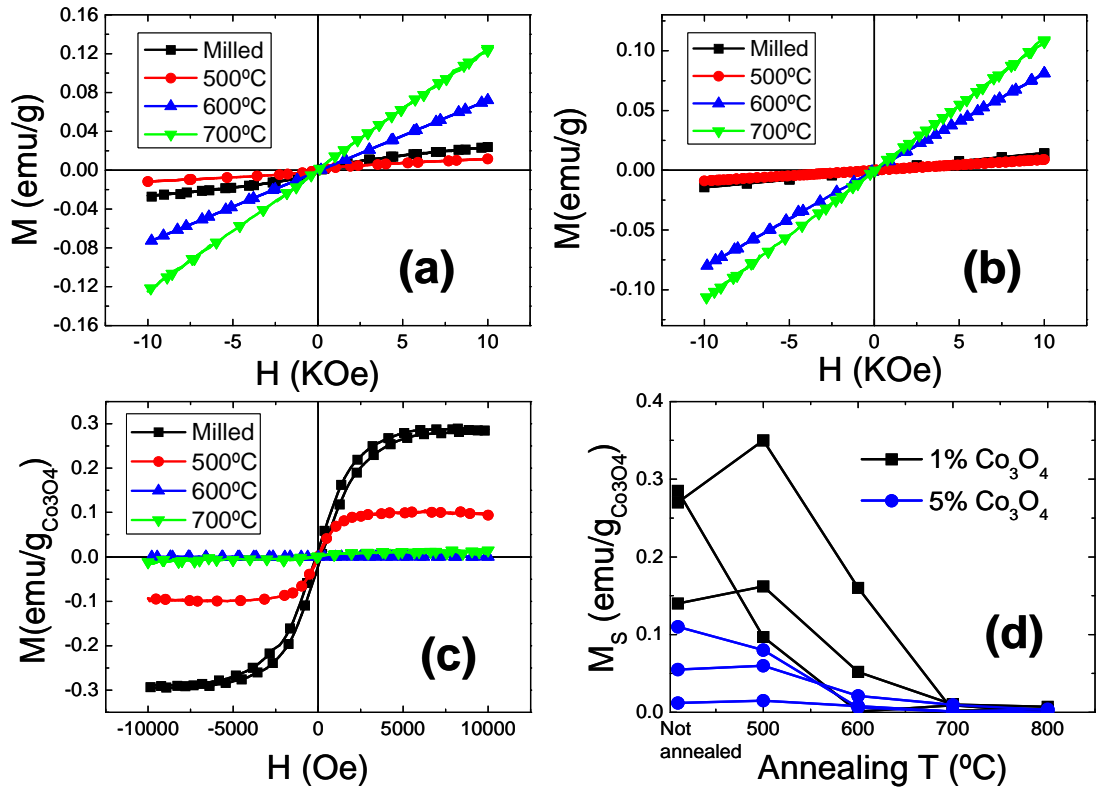

**Figure 4.** (Color online) Magnetization curves for 99% $TiO_2$ - 1% $Co_3O_4$ samples annealed at different temperatures for (a) A-$TiO_2$ and (b) R-$TiO_2$. (c) Curves for the samples with $TiO_2$ in anatase structure after subtraction of a paramagnetic (linear) component. (d) Saturation magnetization of the ferromagnetic component for the samples prepared with A-$TiO_2$ with 1% and 5% wt of $Co_3O_4$.

Despite the diamagnetic character of $TiO_2$ and the antiferromagnetic behavior of $Co_3O_4$ and $CoTiO_3$ (with Neel temperatures of 42K and 37K respectively) we found a ferromagnetic contribution at room temperature for the samples with $Co_3O_4$ and A-$TiO_2$ milled and after annealing at low 500°C and 600°C (figure 4c); this contribution not observed in the R-$TiO_2$ samples in all range of temperatures.

The experiment was repeated three different times at different laboratories, using oxides from different suppliers with similar specifications and different experimental setups. While the quantitative results are different (see figure 4d), in all three cases we found the same effect: Ferromagnetism at room temperature is only observed when using A-$TiO_2$ structure and after annealing at low temperature (below 700°C). Quantitative differences appear in the maximum value of $M_S$, which ranges from $3.6 \cdot 10^{-1}$ to $1 \cdot 10^{-1}$ emu/$g_{Co3O4}$. The highest $M_S$ in the presented data appears after milling, while for the other two sets it is found after 500°C annealing (figure 4d). For the samples with 5% of $Co_3O_4$, the effect also exists, although the value of $M_S$ is fairly smaller than for those samples with 1%$Co_3O_4$. The paramagnetic component, on the other hand, is found to be basically the same for the samples annealed at the same temperature but prepared at different laboratories.

The appearance of room temperature magnetism in $TiO_2$ based materials containing cobalt only for the anatase structure is in agreement with most of the experimental results published for the Ti-O-Co system [16,17,18,19,20] while scarce experiments are found for R-$TiO_2$ [21].

However, the appearance of magnetism just after a short milling is surprising as no diffusion or strong interaction is expected at this stage. Such behavior indicates some interaction between $TiO_2$ and $Co_3O_4$ just after milling. The weak ferromagnetic signal suggests that only a small part of the material exhibits ferromagnetic behavior, which is consistent with the non ferromagnetic behavior of $TiO_2$ and $Co_3O_4$, the only phases observed by XRD in the ferromagnetic samples. The magnetic signal could then arises at the interfaces between both oxides, that are modified in the first stages of the reaction, and are not detected by XRD since they represents a very small fraction of the whole material.

In order to confirm that this early interaction leads to a modification of the electronic configuration of the oxides, we performed optical measurements. Optical reflection spectra for the samples with 5% of $Co_3O_4$ after the different thermal treatments are presented in figure 5. Results for the samples containing 1% of $Co_3O_4$ were similar (although the features arising from the $Co_3O_4$ are not so clearly resolved).

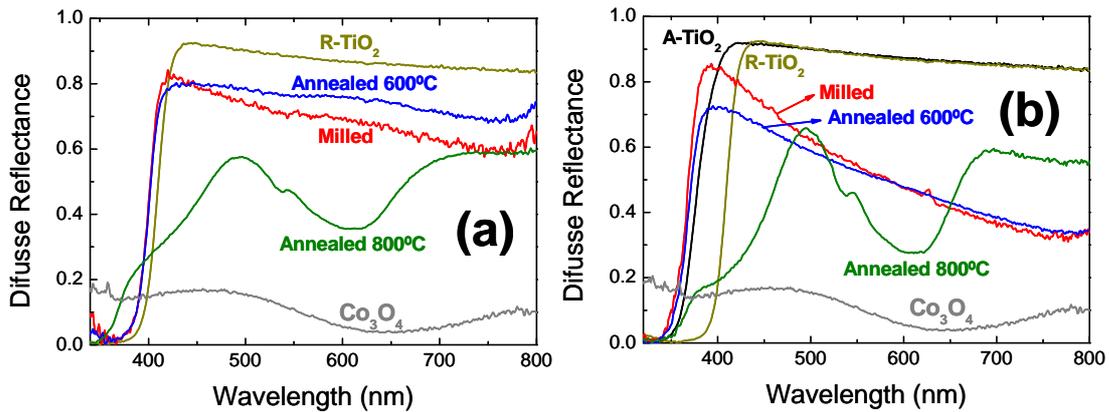

**Figure 5**. (Color online) Diffuse Reflectance spectra for the samples containing 5% $Co_3O_4$ prepared with (a) rutile and (b) anatase $TiO_2$. Spectra from R-$TiO_2$, A-$TiO_2$ and $Co_3O_4$ are also presented for comparison purposes.

Both A-$TiO_2$ and R-$TiO_2$ samples exhibit slight shifts in the edge after milling. This process acts as a very efficiency homogenizer procedure that favors the appearance of $Co_3O_4$-$TiO_2$ particle contacts during the drying step because of Van der Wall forces (figure 1). It is well known that the optical processes are very sensitive to surface effects, which may modify electronic levels or introduce new ones that can account for the observed shifts.

The samples exhibiting ferromagnetic behavior (those prepared with anatase and mixed and low temperature annealed) exhibit a maximum in the reflectance at the edge (390-400 nm) which is absent in all the other samples. This maximum is the combination of the $TiO_2$ edge and a broad absorption band in the green-red par of the spectrum. Actually, the magnetic samples are the only ones exhibiting a blue coloration as the pictures in the supplementary info (figure S1) illustrates. It is particularly noteworthy that the equivalent samples prepared with R-TiO$_2$ show grey color.

Blue color is characteristic of compounds containing Co$^{+2}$ atoms in octahedral positions [22]. However, Co$_3$O$_4$ has spinel structure with the Co$^{+2}$ ions in thetraedral (T) positions and the Co$^{+3}$ ions in octahedral (O) ones [14]. Thus, the blue color could be explained by a reduction of the Co$^{+3}$ ions in O positions to Co$^{+2}$ due to the interaction with the A-TiO$_2$. In order to check this hypothesis, we performed X-ray absorption spectroscopy in the Co K-edge.

Figure 6 presents the XAS spectra at the Co K-edge of pure Co$_3$O$_4$ and samples 95%TiO$_2$-5%Co$_3$O$_4$ samples prepared with A-TiO$_2$ and R-TiO$_2$.

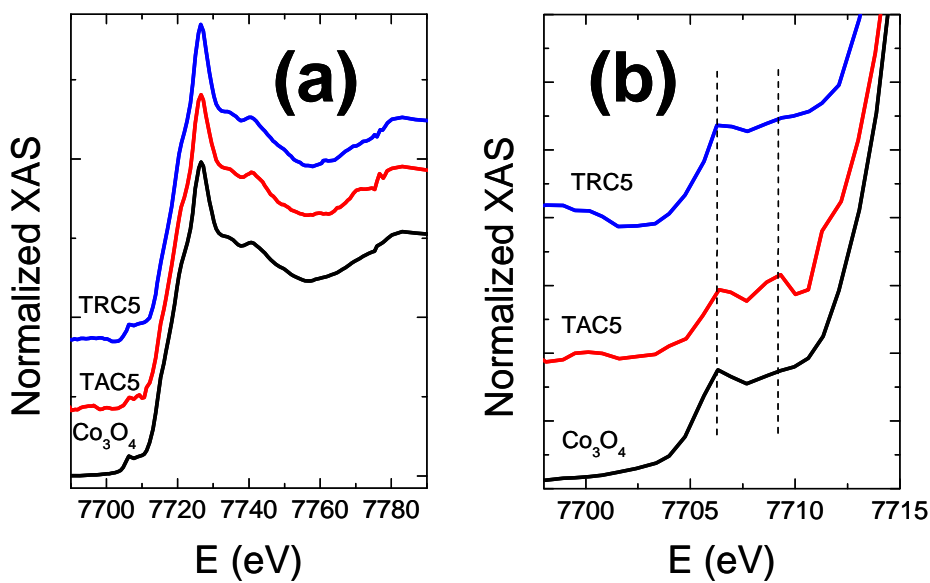

**Figure 6.** (a) (Color online) XANES spectra at the Co K-edge for pure Co3O4 and samples 95%TiO2-5%Co3O4 samples prepared with A-TiO$_2$ (TAC5) and R-TiO$_2$ (TAC5) . (b) Detail of the pre-edge region.

The spectra (figure 6a) are very similar as expected since interaction between both oxides is limited to a narrow surface region. No clear shift of the edge is found. However, the detail of the pre-peak shown in figure 6b evidences some changes. The feature in the $Co_3O_4$ spectra at about 7707 eV corresponds to a bound 1s to 3d transition of tetrahedrally coordinated metal ions [23,24] which is forbidden for positions with a center of inversion as octahedral positions. Therefore, for $Co_3O_4$, this peak is characteristic of $Co^{+2}$ in tetrahedral positions. The pre-peak feature is very similar for the sample prepared with R-$TiO_2$ confirming the weak interaction between $TiO_2$ and $Co_3O_4$ after milling in this sample. On the contrary, for the sample with A-$TiO_2$, the peak is reduced and splits into two, corroborating the modification of the Co electronic structure. The appearance of a new peak at about 7709 eV has been ascribed to Co in octahedral position with a lower oxidation state [25], which agrees with results of the optical spectroscopy and confirms the surface $Co^{+3} \rightarrow Co^{+2}$ reduction in these sample. In our case, the vicinity of Ti atoms may break the octahedral symmetry inducing a distortion that can allow the presence of the pre-peak. Calculations (not presented here) of the XANES spectrum for a $TiO_2/Co_3O_4$ interface with $Co^{+2}$ in octahedral positions, shown a double pre-peak at the same positions than those observed experimentally. A similar behavior was observed for the samples annealed at 500°C and 600°C that exhibit also ferromagnetic signals (see supplementary info figure S2). These results confirm that the early interaction of the $Co_3O_4$ with the A-$TiO_2$ induces the reduction of the $Co^{+3}$ in O positions to $Co^{+2}$, while this effect is not observed in the samples prepared with R-$TiO_2$. Concerning the origin of the interaction, a electrochemical solid state reaction taking place at the oxides surface has been proposed [26].

Optical and XAS spectroscopy studies point on the Co reduction as the origin of the ferromagnetic signal. As indicated above, $Co_3O_4$ presents the spinel structure as most of the ferrimagnetic materials, like $Fe_3O_4$ or $NiCo_2O_4$, with the $Co^{+2}$ ions in T positions and the $Co^{+3}$ ions in O ones [14]. For this structure, the predominant interaction is the antiferromagnetic super-exchange between T and O positions, while interaction T-T and O-O are also antiferromagnetic but much weaker than the T-O one. Consequently, there are two magnetic sublattices corresponding to T and O position with antiparallel orientation. Therefore these spinels exhibit ferrimagnetic behavior with high order temperature ($T_C$=858 K for $Fe_3O_4$, or 673 K for $NiCo_2O_4$). However, for $Co_3O_4$, $Co^{+3}$ ions in O positions present no magnetic moment due to the large splitting of the 3d orbital in this symmetry [14]. Hence, in $Co_3O_4$ only $Co^{+2}$ in T position holds a magnetic moment and the T-T weak antiferromagnetic interaction is the dominant, so the system present antiferromagnetic behavior with a much lower order temperature of $T_N$=42 K. However, for $Co^{+2}$ atoms in an octahedral field, the orbital splitting is quite small and a $Co^{+2}$ atom in this symmetry should hold a magnetic moment [14]. For a region of the crystal where $Co^{+3}$ in O positions are reduced to $Co^{+2}$ there should be magnetic moments in both O and T positions with partially filled $t_{2g}$ orbitals. In this situation, the T-O antiferromagnetic interaction would be the dominant one and the system should present a behavior similar to that of $Fe_3O_4$, that is, ferrimagnetism with high order temperature (due to the strong interaction T-O). Therefore, the observed weak ferromagnetic signal can be explained by the reduction of $Co^{+3}$ in O positions to $Co^{+2}$ demonstrated by XANES and optical spectra, and no new magnetic ordering mechanisms are required to account for it. However, the effect in $Co_3O_4$ will be restricted to very small surface regions; the spinel structure of $Co_3O_4$ is unstable if a large fraction of $Co^{+3}$ is reduced to $Co^{+2}$ promoting the transformation to CoO which is antiferromagnetic. Thus, it is not possible to obtain a bulk ferrimagnetic material nor a material with uniform magnetic properties based on this effect.

Surface is the most sensitive part of the materials to any kind of treatment or interaction. Thus, nominally identical oxides from different suppliers can present identical bulk properties but fairly different surfaces. Properties depending on the surface and its reactivity in the first stages can be completely different, as we found for the three sets of samples analyzed here. This experimental fact could account for the discrepancies in results from different groups that use powder nominally identical with the same bulk structure but different surfaces due to different origin. This idea is also supported by recent results [27] that show that a pre-calcination (just at 400°C) of ZnO alters its surface electronic structure and consequently its reactivity.

## Conclusions

In summary, we have demonstrated that a weak interaction between Anatase-$TiO_2$ and $Co_3O_4$ surface induces a surface reduction of $Co^{+3}$ atoms in octahedral positions to $Co^{+2}$ generating a weak ferromagnetic signal at room temperature. This magnetic signal can be explained as due to super-exchange interactions (as the situation is similar to that in $Fe_3O_4$) and no new magnetic ordering mechanisms are required to account for them. This demonstrates the possibility to observe weak ferromagnetic signals in samples containing Ti, O and Co that must be considered when studying the magnetic properties of Co:$TiO_2$ diluted magnetic semiconductors and similar materials.

## Acknowledgements:
This work was supported by the Spanish Council for Scientific Research through the projects CSIC 2006-50F0122 and CSIC 2007-50I015 and Spanish Ministry of Science and Education through the projects MAT2007-66845-C02-01 and FIS-2008-06249. We acknowledge the European Synchrotron Radiation Facility for provision of synchrotron radiation facilities and we would like to thank the SpLine CRG beamline staff for assistance during X-Ray absorption experiments.

# Supplementary Info

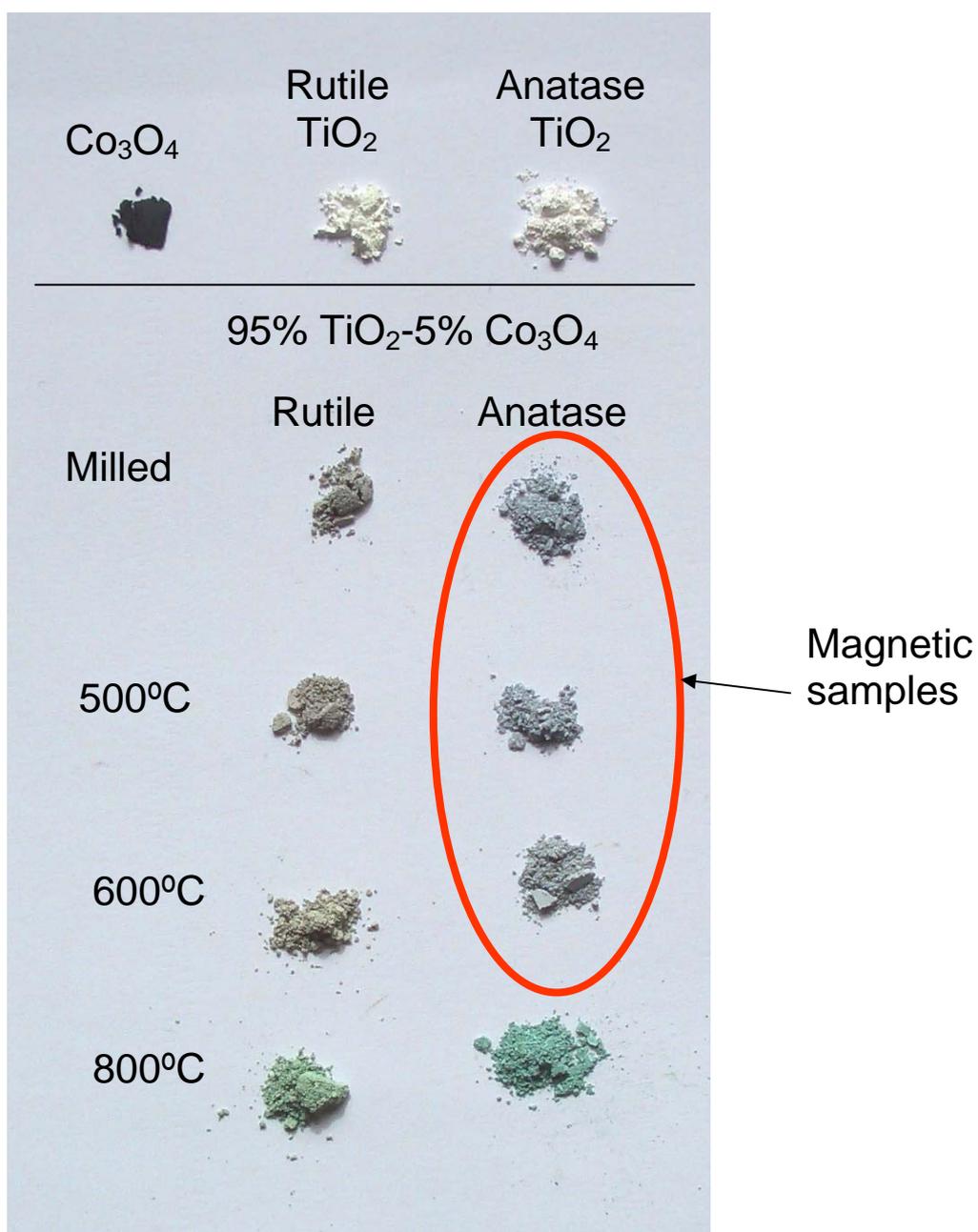

Figure S1. Photograph of the $Co_3O_4$, R-TiO$_2$ , A-TiO$_2$ and samples prepred with these powders. The showing ferromagnetic signals are those indicated by the red circle which exhibit bluish coloration.

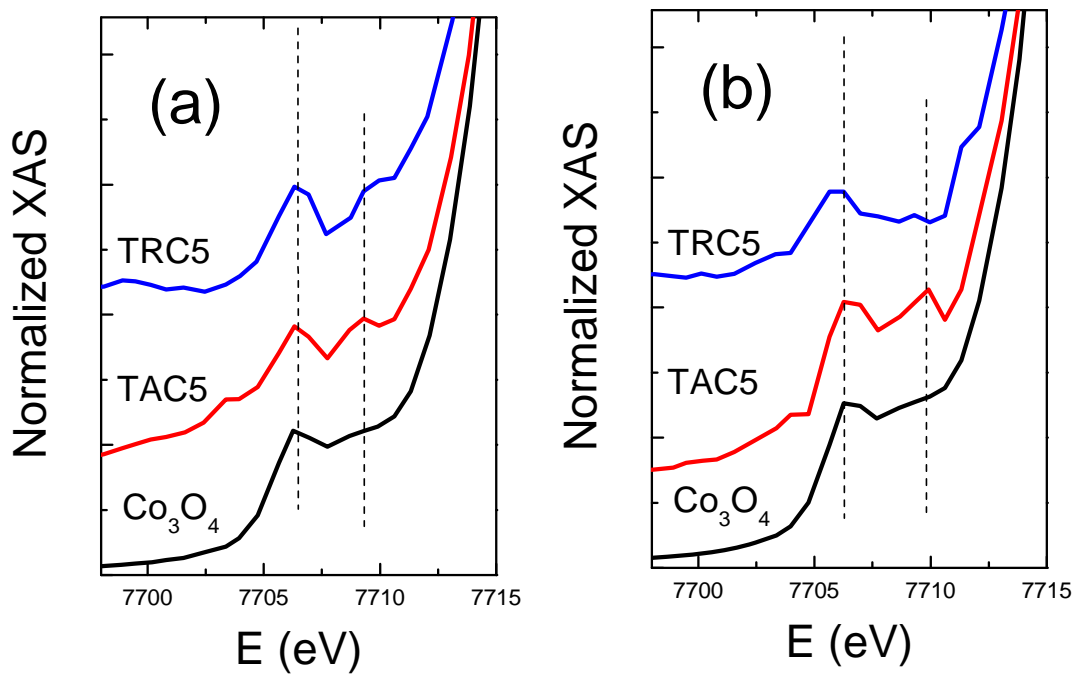

Figure S2. Detail of the XAS spectra at the Co K-edge for the the $Co_3O_4$, and 95%$TiO_2$-5%$Co_3O_4$ prepared with anatase $TiO_2$ (sample TAC5) and rutile $TiO_2$ (samples (TRC5) after annealing in air at (a) 500°C and (b) 600°C.